\documentclass[12pt]{article}
\newcommand{\R}{\mathcal{R}}
\usepackage{amsfonts}
\usepackage{amsmath}
\usepackage{xcolor}
\usepackage{comment}
\usepackage{graphicx}
\usepackage{booktabs}

\title{The challenges of modeling and forecasting the spread of COVID-19}

\author{Andrea L. Bertozzi\and Elisa Franco \and George Mohler \and Martin B. Short \and Daniel Sledge}

\begin{document}

\maketitle
\begin{abstract}
  We present three data driven model-types for COVID-19 with a minimal number of parameters to provide insights into the spread of the disease that may be used for developing policy responses.  The first is exponential growth, widely studied in analysis of early-time data.  The second is a self-exciting branching process model which includes a delay in transmission and recovery.  It allows for meaningful fit to early time stochastic data.  The third is the well-known Susceptible-Infected-Resistant (SIR) model and its cousin, SEIR, with an "Exposed" component.  All three models are related quantitatively, and the SIR model is used to illustrate the potential effects of short-term distancing measures in the United States. 
\end{abstract}



The world is in the midst of an ongoing pandemic, caused by the emergence of a novel coronavirus.  Pharmaceutical interventions such as vaccination and anti-viral drugs are not currently available. In the short run, addressing the COVID-19 outbreak will depend critically on the successful implementation of public health measures including social distancing, workplace modifications, disease surveillance, contact tracing, isolation, and quarantine.

On March 16th, Imperial College London released a report \cite{Ferguson20COVID} predicting dire consequences if the US and UK did not swiftly take action. In response, in both the US and the UK, governments responded by implementing more stringent social distancing regulations \cite{Lander20}. We now have substantially more data, as well as the benefit of analyses performed by scientists and researchers across the world \cite{Imai,li2020simulating,Riou20,Perkins,EarlyR0,Tindale20,hellewell2020feasibility,WuLancet}. Nonetheless, modeling and forecasting the spread of COVID-19 remains a challenge.

Here, we present three basic models of disease transmission that can be fit to data provided by the Imperial College report and to data coming out of different cities and countries. While the Imperial college study employed an agent-based method (one that simulates individuals getting sick and recovering through contacts with other individuals in the population), we present three macroscopic models: (a) exponential growth; (b) self-exciting branching process; and (c) the SIR compartment model. These models have been chosen for their simplicity, minimal number of parameters, and for their ability to describe regional-scale aspects of the pandemic.

Because these models are parsimonious, they are particularly well-suited to isolating key features of the pandemic and to developing policy-relevant insights. We order them according to their usefulness at different stages of the pandemic - exponential growth for the initial stage, self-exciting branching process when one is still analyzing individual count data going into the development of the pandemic, and a macroscopic mean-field model going into the peak of the disease.

From a public policy perspective, these models highlight the significance of fully-implemented and sustained social distancing measures. Put in place at an early stage, distancing measures that reduce the virus's reproduction number -- the expected number of individuals that an infected person will spread the disease to -- may allow much-needed time for the development of pharmaceutical interventions, or potentially stop the spread entirely. By slowing the speed of transmission, such measures may also reduce the strain on health care systems and allow for higher-quality treatment for those who become infected. The models presented here demonstrate that relaxing these measures in the absence of pharmaceutical interventions prior to the outbreak's true end will allow the pandemic to reemerge. Where this takes places, social distancing efforts that appear to have succeeded in the short term will have little impact on the total number of infections expected over the course of the pandemic. 

This work is intended for a broad science-educated population, and includes explanations that will allow scientific researchers to assist with public health measures. We also present examples of forecasts for viral transmission in the United States. The results of these models differ depending on whether the data employed cover infected patient counts or mortality. In addition, many aspects of disease spread, such as incubation periods, fraction of asymptomatic but contagious individuals, seasonal effects, and the time between severe illness and death are not considered here.  

\section{Results}
\subsection{Exponential Growth}
Epidemics naturally exhibit exponential behavior in the early stages of an outbreak, when the number of infections is already substantial but recoveries and deaths are still negligible. If at a given time $t$ there are $I(t)$ infected individuals, 
and $\alpha$ is the rate constant at which they infect others, then at early times (neglecting recovered individuals), 
$I(t)=I_0e^{\alpha t}$. The time it takes to double the number of cumulative infections (doubling time) is a common measure of how fast the contagion spreads: if we start from $\bar{I}$ infections, it takes a time $T_d=\ln{2}/\alpha$ to achieve $2\bar{I}$ infections.

\begin{figure*}
 \begin{center}
\includegraphics[width=.9\linewidth]{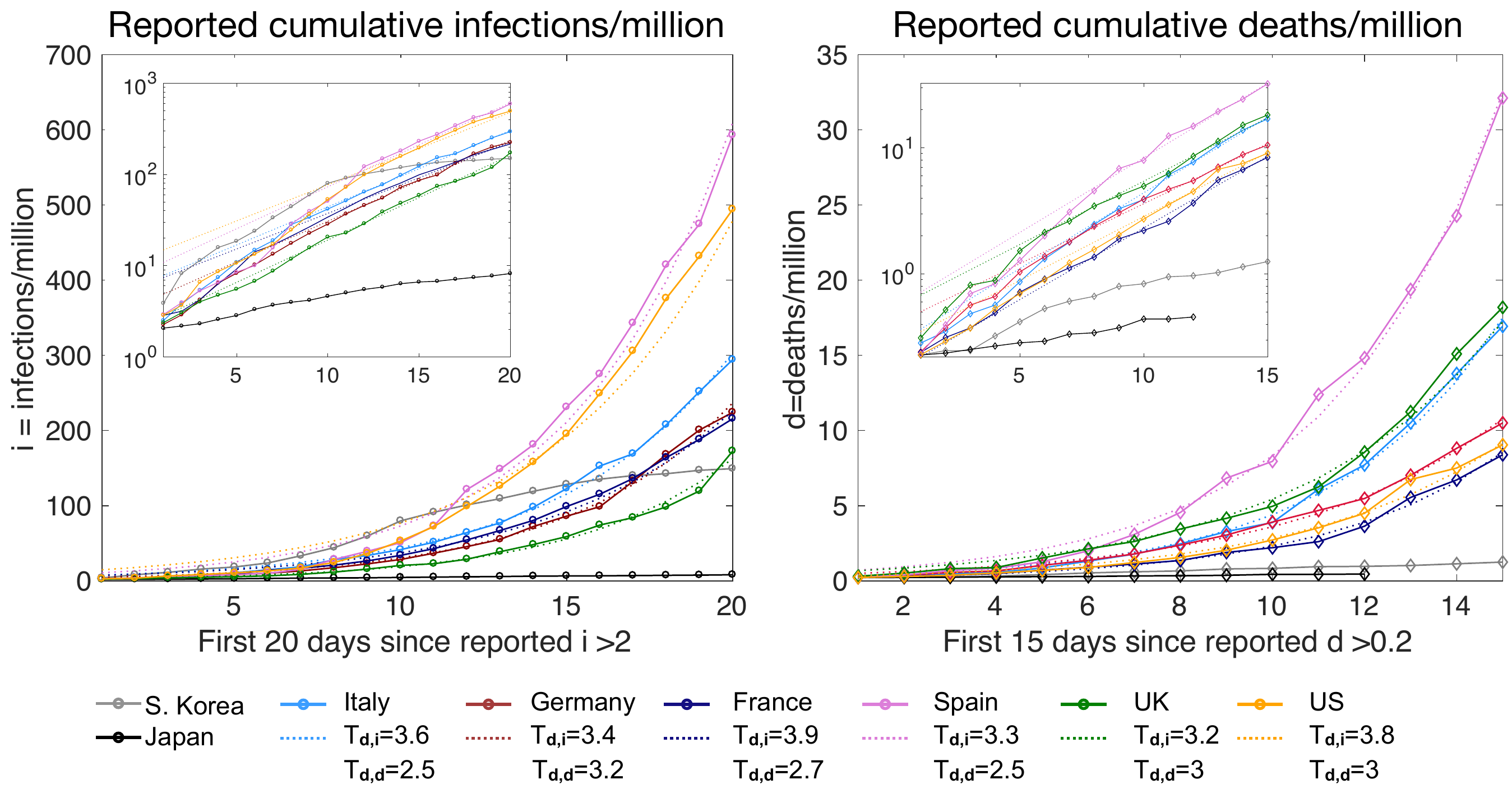}
\includegraphics[width=.9\linewidth]{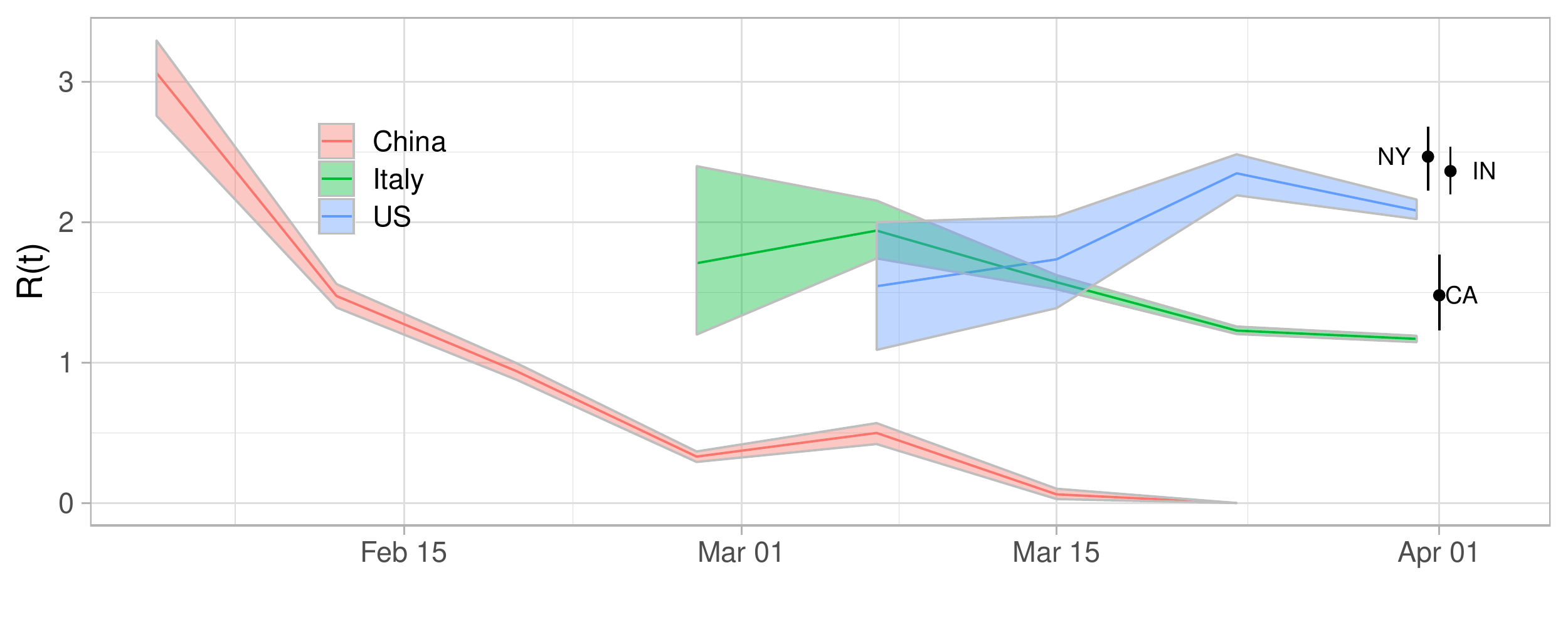}
\caption{\small (a) Exponential model applied to new infection and death data for Italy, Germany, France, Spain, the UK, and the United States, normalized by the total country population (source, WHO). Insets show the same data on a logarithmic scale. Both the normalized infection $i$ and death $d$ data were thresholded to comparable initial conditions for each country; fits are to the first 15-20 days of the epidemic after exceeding the threshold. The fitted doubling time is shown for both infections ($T_{d,i}$) and death ($T_{d,d}$) data. Data from Japan and South Korea are show for comparison and do not exhibit exponential growth. 
(b) Dynamic reproduction number (mean and 95\% confidence interval) of COVID-19 for China, Italy, and the United States estimated from reported deaths \cite{plague} using a non-parametric branching process \cite{HawkesR0COVID}. Current estimates as of April 1, 2020 of the reproduction number in New York, California, and Indiana (confirmed cases used instead of mortality for Indiana). Reproduction numbers of Covid-19 vary in different studies and regions of the world (in addition to over time), but have generally been found to be between 1.5 and 6 \cite{LiuTravelMedicine2020} prior to social distancing.
\label{fig:Exponential}}
 \vspace{-.5cm}
\end{center}
\end{figure*}
For the COVID-19 outbreak, exponential growth is seen in available data from multiple countries (see Figure~\ref{fig:Exponential}),  with remarkably similar estimated doubling times in the early stages of the epidemic. For COVID-19, we expect an exponential growth phase during the first 15-20 days of the outbreak, in the absence of social distancing policies. This estimate is based on patient data from the Wuhan outbreak, which indicate that the average time from illness onset to death or discharge is between 17 and 21 days for hospitalized patients \cite{pan2020time, zhou2020clinical}. Because they are a fraction of infections, deaths initially increase at a similar exponential pace, with some delay relative to the beginning of the outbreak.   These observed doubling time estimates are significantly smaller than early estimates ($\sim$7 days) obtained using data collected in Wuhan from field investigations \cite{NEJMWuhan}. 

\subsection{Self-exciting point processes}

A branching point process \cite{Rpackage,farrington2003branching,RickAISM19} can also model the rate of infections over time. Point processes are easily fit to data and allow for parametric or nonparametric estimation of the reproduction number and transmission time scale.  They also allow for estimation of the probability of extinction at early stages of an epidemic.  These models have been used for various social interactions including spread of Ebola \cite{EbolaHawkes}, retaliatory gang crimes \cite{Stomakhin},
and email traffic \cite{Fox,Zipkin}.
The intensity (rate) of infections can be modeled as
\begin{equation}
    \lambda(t) = \mu+\sum_{t_i<t}
    \R(t_i) w(t-t_i)
    \label{eq1}
\end{equation}
where $t$ is the current time and $t_i$ are the times of previous infection incidents.  Here the dimensionless reproduction number, $\R (t)$, evolves in time \cite{cauchemez2006real,wallinga2004different,forsberg2008likelihood,Obadia12,HawkesR0COVID} to reflect changes in disease reproduction in response to public health initiatives (e.g. school closings, social distancing, closures of non-essential businesses).   The distribution of inter-event times $w(t_i-t_j)$ is typically modeled using a gamma or Weibull distribution \cite{Obadia12,cowling2010effective,hellewell2020feasibility}; we choose Weibull with shape parameter $k$ and scale parameter $b$.  Finally, the parameter $\mu$ allows for exogenous infection cases.  Given \eqref{eq1}, the quantity
\begin{equation}
 p_{ij}=\R(t_j)w(t_i-t_j)/\lambda(t_i)
 \label{eq2}
\end{equation}
gives the probability of secondary infection $i$ having been caused by primary infection $j$. The point process in \eqref{eq1} can be viewed as an approximation to the common SIR model of infectious diseases (described later) during the initial phase of an epidemic when the total number of infections is small compared to the overall population size \cite{SIRHawkes}.

Figure~\ref{fig:Exponential}(b) shows the estimated dynamic reproduction number \cite{you2020estimation,riley2003transmission} of COVID-19 in China, Italy, and the U.S. from late January, 2020 to early April, 2020.  The branching point process is fit to mortality data \cite{plague} using an expectation-maximization algorithm \cite{HawkesR0COVID}.  Public health measures undertaken in China appear to have reduced  $\R (t)$ to below the self-sustaining level of $\R=1$ by the middle of February.  In Italy, public health measures brought the local value of $\R (t)$ down; as of early April, however, it remained above $\R=1$. Currently, the estimated reproduction number in the U.S. as a whole is around $2.5$. The reproduction number, however, varies notably by location.

The branching point process model can also be adapted to capture the long-term evolution of the pandemic by incorporating a pre-factor that accounts for the eventual decrease in the number of susceptible individuals  \cite{SIRHawkes}:
\begin{equation}
\lambda^h(t)=(1-N_t/N) (\mu +\sum_{t>t_i}\R(t_i)w(t-t_i)).
\label{eq:SIRHawkes}
\end{equation}
Here $N_t$ is the cumulative number of infections as of time $t$ and $N$ is the total population size.  This version of the branching process model, referred to as HawkesN, represents a stochastic version of the SIR model (described below); with large $\R$, the results of HawkesN are essentially deterministic. When projecting, we use our estimated $\R(t_i)$ at the last known point for all times going forward.  Since the $N_t$ term is the number of infections, if our estimates for $\R(t_i)$ are based on mortality numbers, we must also choose a mortality rate to interpolate between the two counts; though estimated rates at this time seem to vary significantly, we choose 1\% as a plausible baseline \cite{Verity}. Alternatively, we also create forecasts for three US states based on fits to reported case data (see Table \ref{fits}).


\begin{table*}
\includegraphics[width=0.8\textwidth]{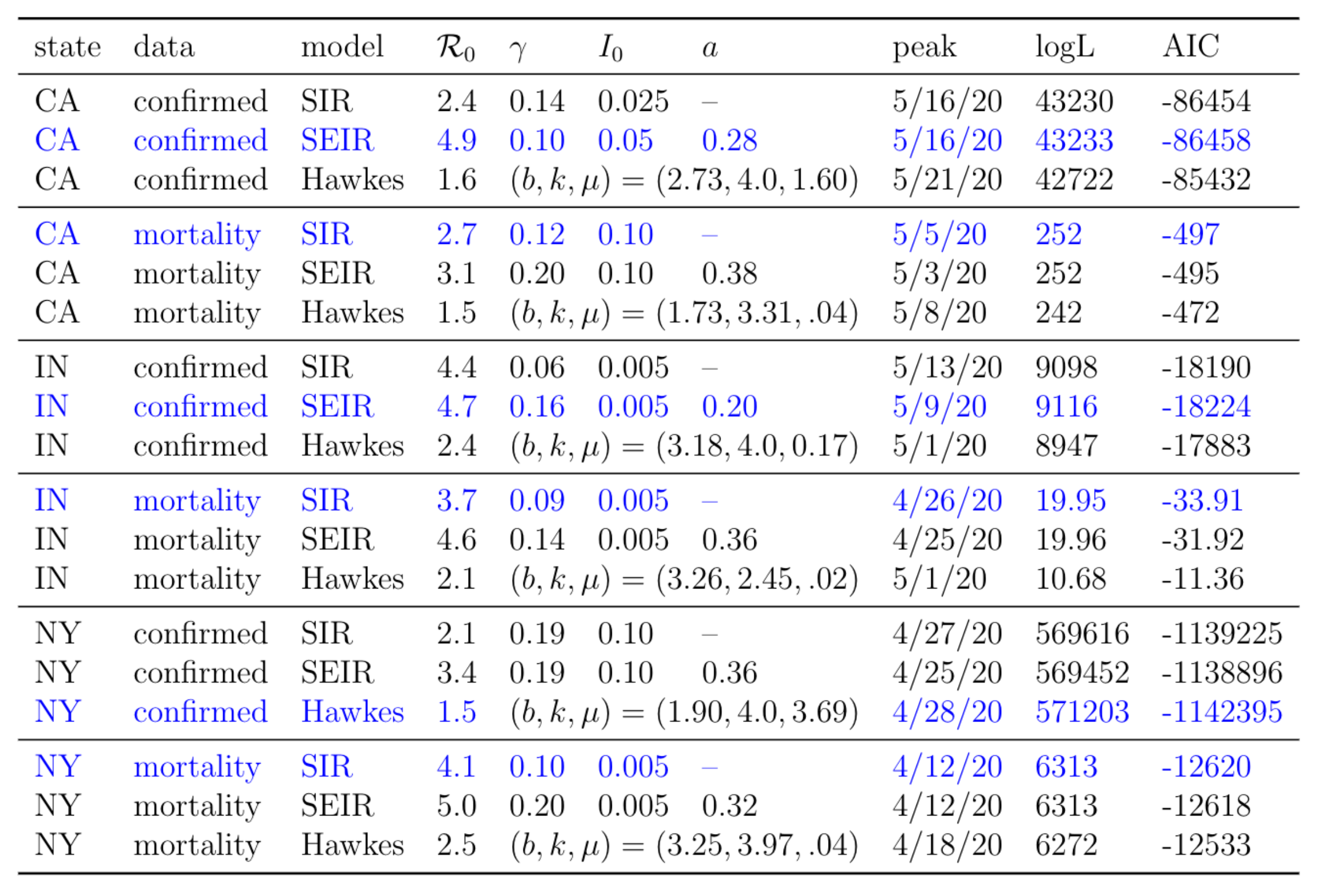}\includegraphics[width=0.3\textwidth, height=75mm]{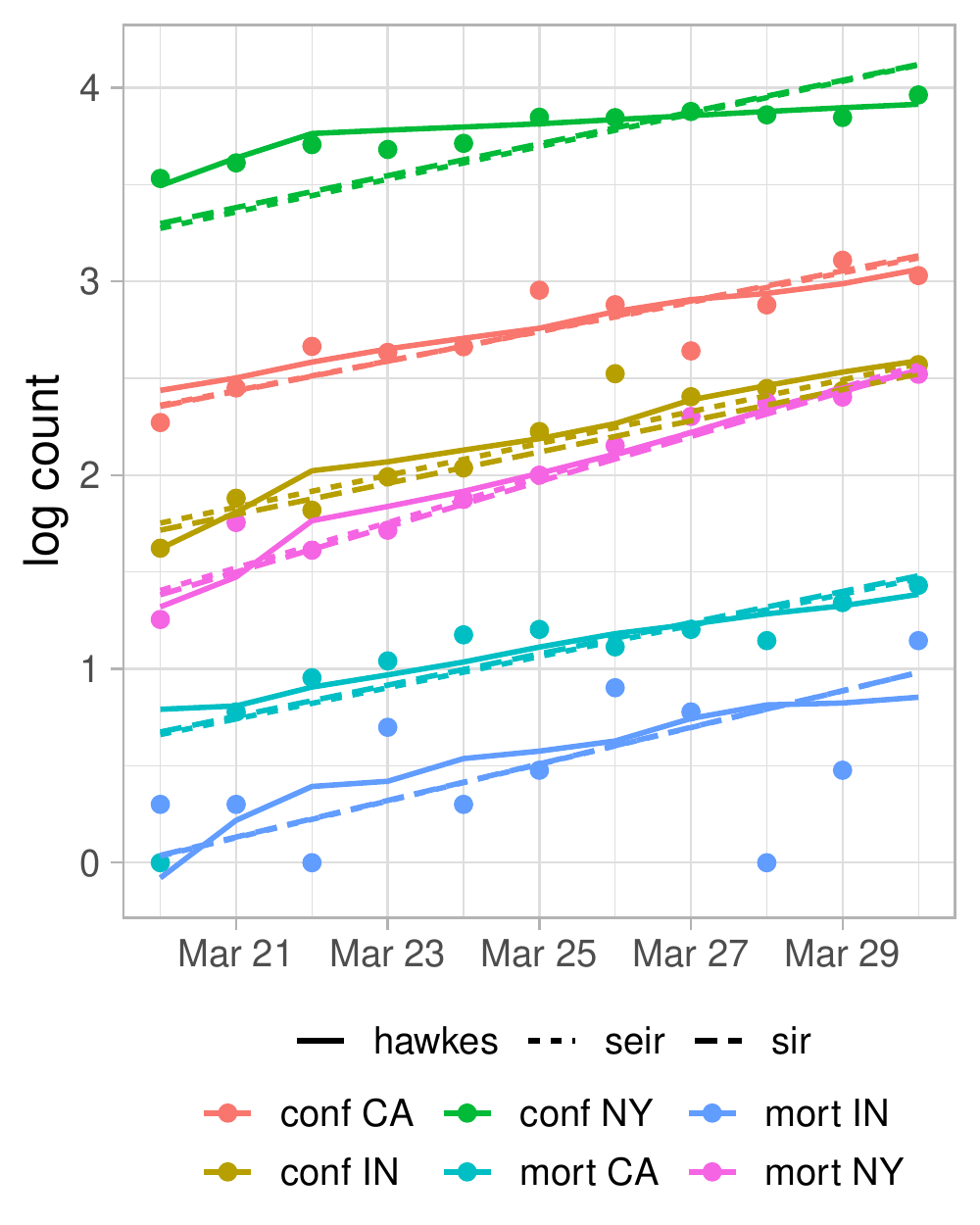}
\caption{\small (Left) Fit of data from California, Indiana, and New York States to three different models, SIR, SEIR, and HawkesN, using Poisson regression. The log-likelihood and the Akaike Information Criteria \cite{AIC} are shown.  The blue lettering corresponds to the lowest AIC value. The Hawkes process parameters include a Weibull shape $k$ and scale $b$ for $w(t)$, along with the exogenous rate $\mu$.  Left shows parameters from the fit and the projected date for the peak in new cases for each of these datasets.  For each state, we run the fit on both confirmed case data and mortality data, taken from \cite{plague}.  (Right) Shown are the actual data points compared to the fitted curves.\label{fits}}
\end{table*}
\subsection{Compartmental Models}

The SIR model \cite{Wiki-compartment,SIR27} describes a classic “compartmental” model with Susceptible-Infected-Resistant population groups. A related model, SEIR, including an “Exposed” compartment, was shown to fit historical death record data from the 1918 Influenza epidemic \cite{Bootsma7588}, during which governments implemented extensive social distancing measures, including bans on public events, school closures, and quarantine and isolation measures. The SIR model can be fit to the predictions made in \cite{Ferguson20COVID} for agent based simulations of the United States.
The SIR model assumes a population of size $N$ where $S$ is the total number of susceptible individuals, $I$ is the number of infected individuals, and $R$ is resistant.  For simplicity of modeling, we view deaths as a subset of resistant individuals and deaths can be estimated from the dynamics of $R$; this is reasonable for a disease with a relatively small death rate. We also assume a short enough timescale during which resistance does not degrade sufficiently.  We do not yet have sufficient data to know what that time is although it is reasonable to consider resistance to last among the general population for several months.

The SIR model equations are
\begin{equation}
\frac{dS}{dt}=-\beta\frac{IS}{N}, \qquad \frac{dI}{dt}=\beta\frac{IS}{N}-\gamma I,  \qquad   \frac{dR}{dt}=\gamma I,     \label{SIR}
\end{equation} $\R_0=\beta/\gamma$ .
Here $\beta$ is the transmission rate constant, $\gamma$ is the recovery rate constant, and $\R_0$ is the reproduction number. One integrates \eqref{SIR} forward in time from an initial value of $S$, $I$, and $R$ at time zero.   
The SEIR model includes an Exposed category $E$:
\begin{align*}
\frac{dS}{dt}&=-\beta\frac{IS}{N}, \qquad
\frac{dE}{dt}=\beta\frac{IS}{N}-a E, \\
\frac{dI}{dt}&=aE-\gamma I,  \qquad  \frac{dR}{dt}=\gamma I.      
\end{align*}
 Here $a$ is the inverse of the average incubation time.  Both models are fit, using maximum likelihood estimation with a Poisson likelihood, to data for three US States (CA, NY, and IN) \cite{plague}.  The results are shown in Table~\ref{fits} with a comparison to HawkesN.  We use the Akaike information criteria \cite{AIC} to measure model performance for each dataset; it is biased against models with more parameters.  The SEIR model performs better on the Confirmed data for California and New York State, possibly due to the larger amount of data, compared to mortality for which SIR is the best for all three states.  HawkesN performs best for confirmed cases in NY. 
 
 \begin{figure*}
\centering
\includegraphics[width=.6\linewidth]{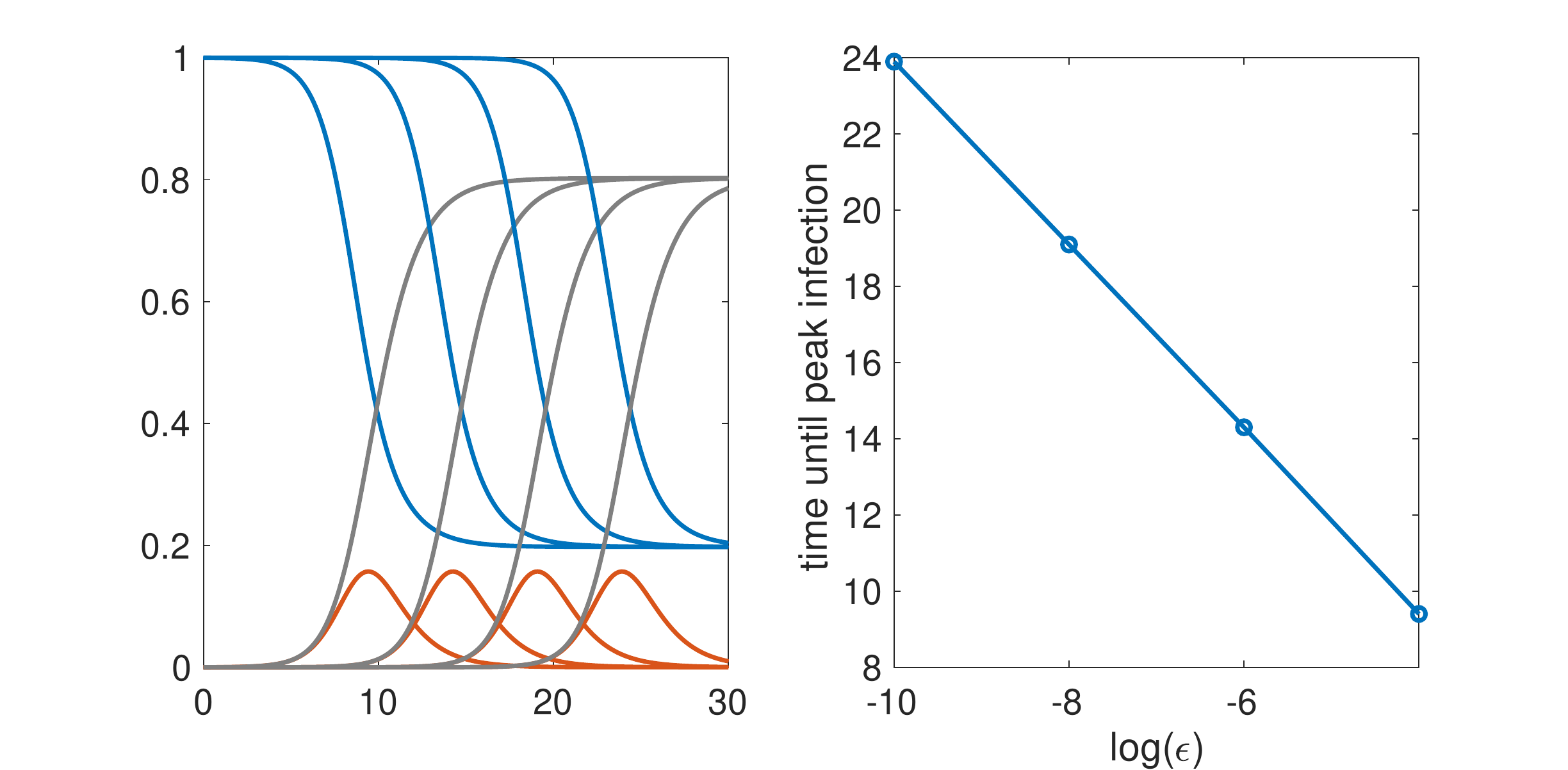}\hspace*{-.7cm}
\includegraphics[width=0.26\linewidth]{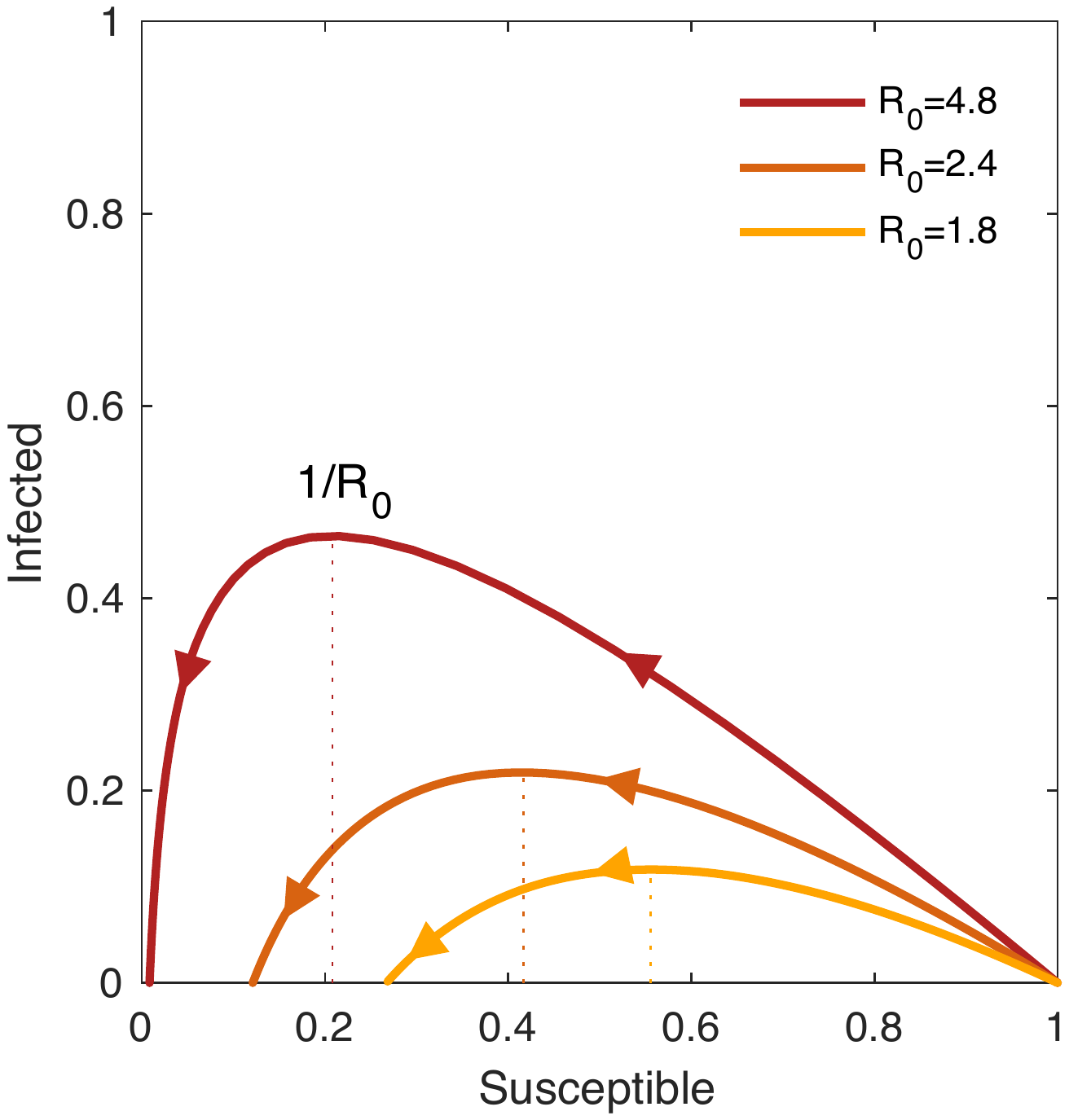}\vspace*{-1cm}
\includegraphics[width=0.7\linewidth]{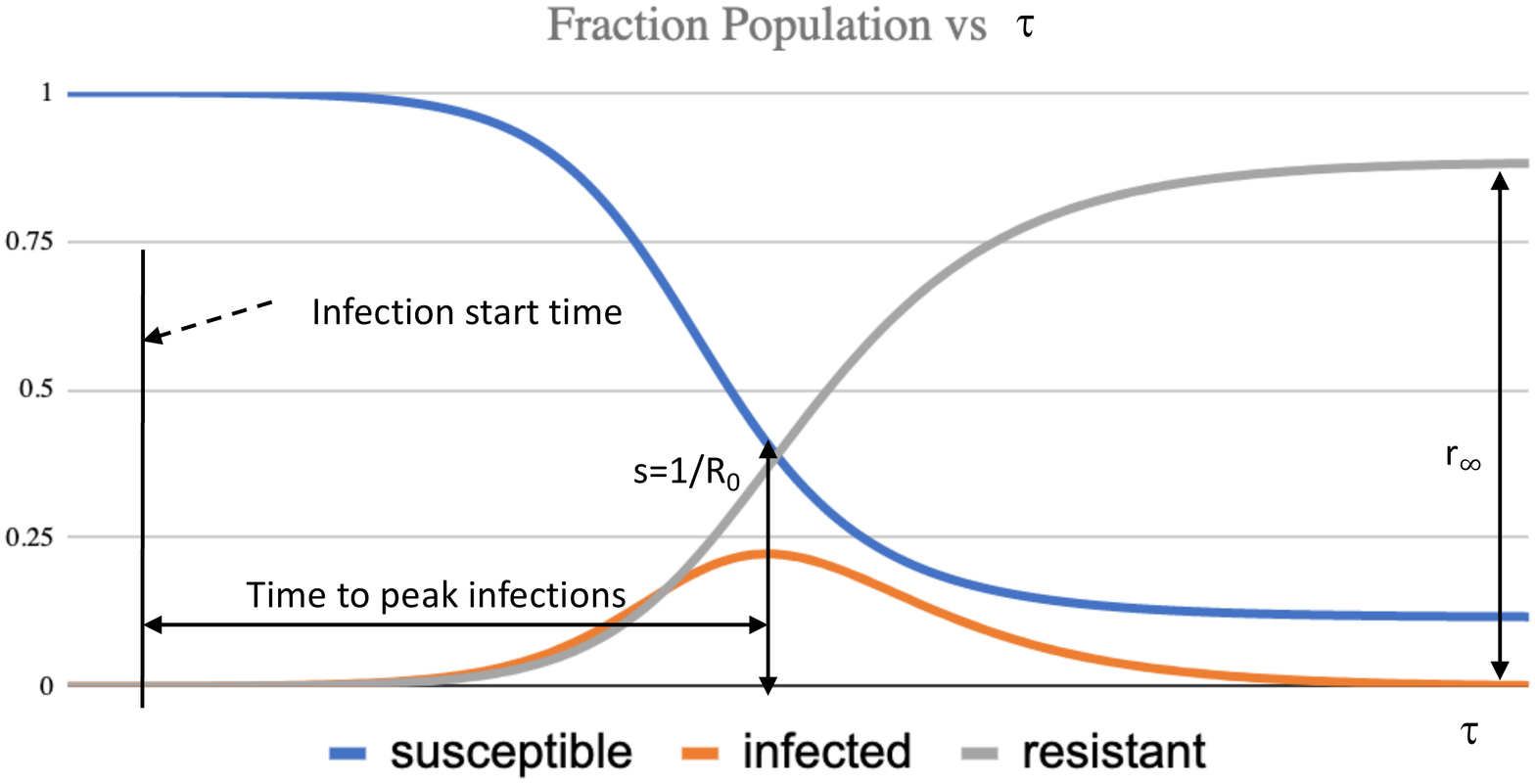}
\caption{\small Solution of dimensionless SIR model (5) with $\R_0=2$. The first panel show the graphs of $s$ (blue), $i$ (orange) and $r$ (grey) on the vertical axis vs. $\tau$ on the horizontal axis, for different $\epsilon$.  The corresponding values of $\epsilon$ from left to right are $10^{-4}$, $10^{-6}$ , $10^{-8}$ , $10^{-10}$.  Middle panel shows the time until peak infections vs $\log(\epsilon)$ for the values shown in the left panel.   This asymptotic tail to the left makes it challenging to fit data to SIR in the early stages.  Top right is a phase diagram for fraction of infected vs. fraction of susceptible with the direction of increasing $\tau$ indicated by arrows, for three different values of $\R_0$.  The bottom panel displays a typical set of SIR solution curves over the course of an epidemic, with important quantities labeled.
\label{fig:shift_plot}}

\vskip -0.5in
\end{figure*}

Dimensionless models are commonly used in physics to understand the role of parameters in the dynamics of the solution (a famous example being the Reynolds number in fluid dynamics). The compartmental models \eqref{SIR} have a dimensionless form. 
  There are two timescales dictated by $\beta$ and $\gamma$, so if time is rescaled by $\gamma$, $\tau=\gamma t$, and $s= S/N$, $i=I/N$, and $r= R/N$ represent fractions of the population in each compartment, then we retain only one dimensionless parameter $\R_0$ that, in conjunction with the initial conditions, completely determines the resulting behavior.  
  There are three timescales in SEIR, thus resulting in a dimensionless equation with two dimensionless parameters.
 For SIR, given an initial population with $r(0)= 0$ and any sufficiently small fraction of initial infected $\epsilon=i(0)$, the shapes of the solution curves $s(\tau), i(\tau), r(\tau)$ do not depend on $\epsilon$, other than exhibiting a time shift that depends logarithmically on $\epsilon$ (Fig, ~\ref{fig:shift_plot}).  This is a universal similarity solution for the SIR model in the limit of small $\epsilon$  (Fig. 3), depending only on $\R_0$. Critically, the height of the peak in $i(t)$ and the total number of resistant/susceptible people by the end of the epidemic are determined by $\R_0$ alone.  But, the sensitivity of the time translation to the parameter $\epsilon$, and the dependence of true time values of the peak on parameter $\gamma$ makes SIR challenging to fit to data at the early stages of an epidemic when Poisson statistics and missing information are prevalent.  
 All of this is important information for public health officials, policymakers, and for political leaders to understand, in terms of the importance of decreasing $\R_0$ for potentially substantial periods of time, explaining why projections of the outbreak can display large variability, and highlighting the need for extensive disease testing within the population to help track the epidemic curve accurately. 

After the surge in infections the model asymptotes to end states in which $r$ approaches the end value $r_{\infty}$  and $s$ approaches $1-r_{\infty}$  and the infected population approaches zero. The value $r_{\infty}$ satisfies a well-known transcendental equation \cite{Miller12,Miller17,Harko14}. A phase diagram for the similarity solutions is shown on Fig. \ref{fig:shift_plot} (right).  The dynamics start in the bottom right corner where $s$ is almost 1 and follow the colored line to terminate on the $i=0$ axis at the value $s_\infty$.  A rigorous derivation of the limiting state under the assumptions here can be found in \cite{Harko14,Miller12,Miller17}.

\begin{figure*}
\centering
\includegraphics[width=.65\linewidth]{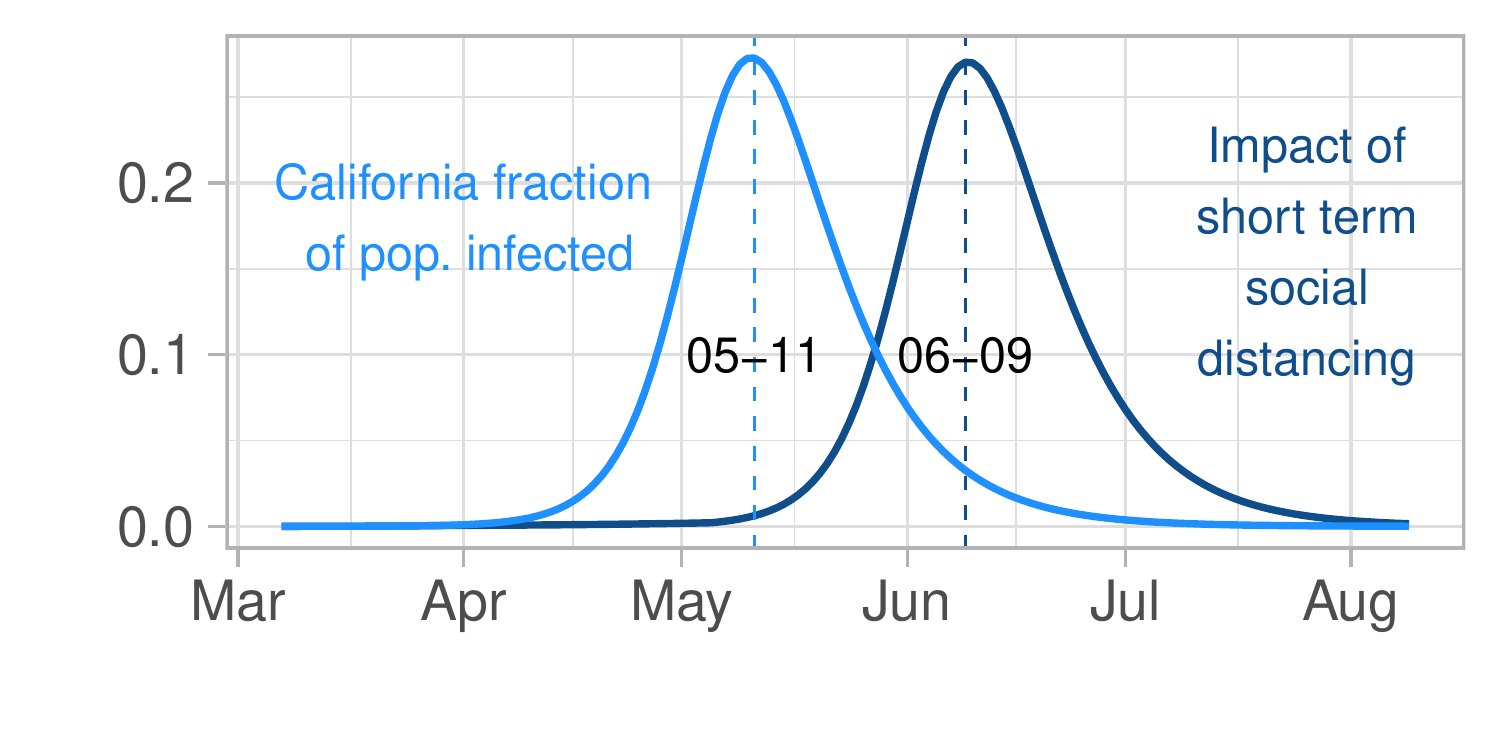}

\includegraphics[width=.65\linewidth]{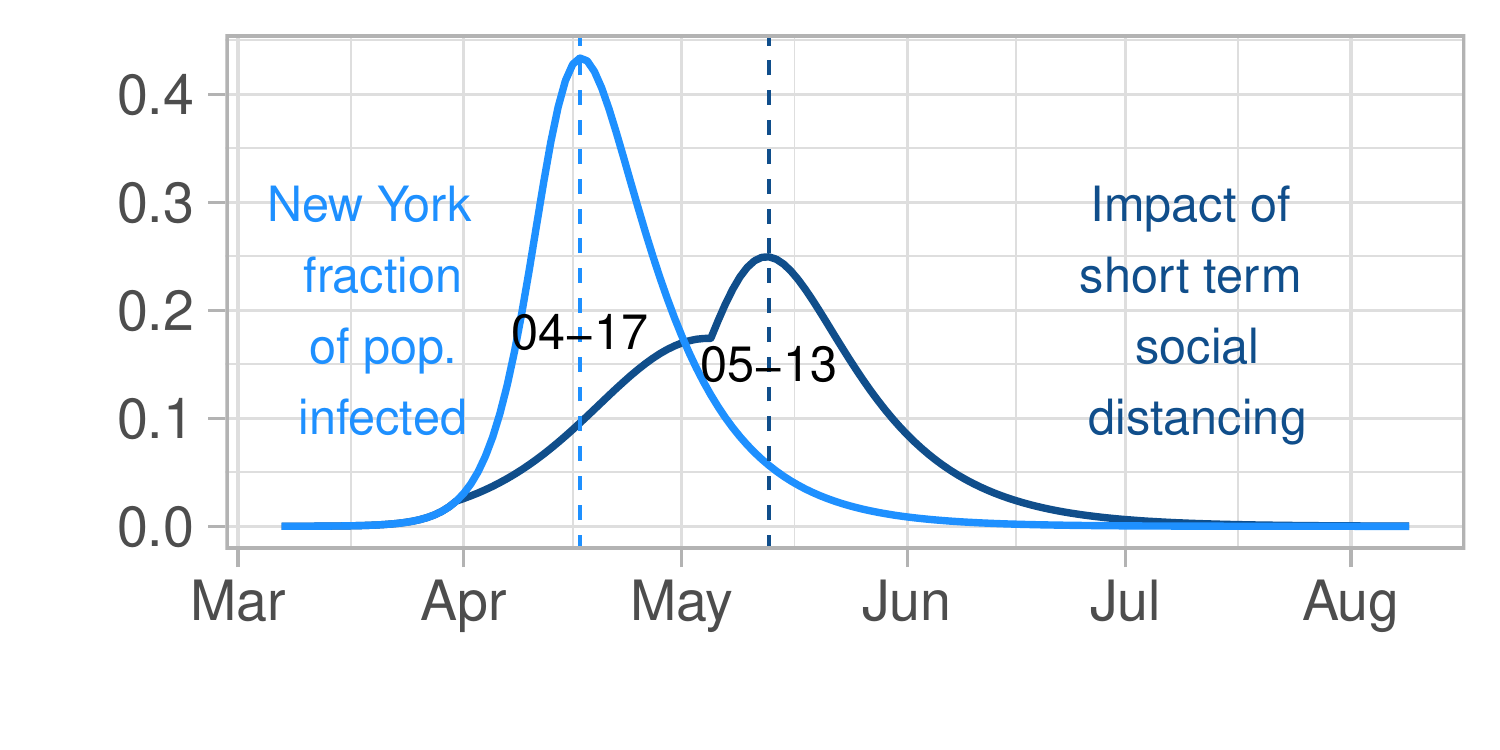}
\vspace*{-.5cm}
\caption{\small Impact of short-term social distancing: fraction of population vs. date.  (Top) California SIR model based on mortality data with parameters from Table \ref{fits} ($\R_0=2.7$,
$\gamma=.12$,
$I_0=.1$).  $\R_0$ is cut in half from March 27 (one week from the start of the California shut down) to May 5 to represent a short term distancing strategy.  (Bottom) New York SIR model with parameters from the Table \ref{fits} ($\R_0=4.1$,
$\gamma=.1$,
$I_0=05$).  We compare the case with no distancing, on the left, to the case with distancing from March 30 (one week from the start of the New York shut down) to May 5.   The distancing measures suppress the curve but are insufficient to fully flatten it.  
}
\label{fig:social_dist}

\end{figure*}

\section*{Discussion}

The analysis presented here illustrates several key points, which can be understood using these parsimonious models.
(a)  The reproduction number $\R$ is highly variable both in time and by location, and this is compounded by distancing measures. These variations can be calculated using a stochastic model and lower $\R$ is crucial for flattening the curve.
(b) Mortality data and confirmed case data have statistics that vary by location and by time depending on testing and on accurate accounting of deaths due to the disease, and can lead to different projected outcomes.
(c) While early control provides time for health providers, it has little effect on the long term outcomes of total infected unless it is sustained. New social protocols may be needed both for the workforce and for society as a whole if we are to avoid both high total levels of infection and a longer term shut down.

Reducing the reproduction number is critical to reducing strain on health care systems, saving lives, and to creating the space for researchers to develop effective pharmaceutical interventions, including a vaccine and anti-viral therapies. While social and economic strains, along with political considerations, may cause policymakers to consider scaling social distancing measures back once shown to be effective, it is critical that leaders at all levels of government remain aware of the dangers of doing so. During the 1918 influenza pandemic, the early relaxation of social distancing measures led to a swift uptick in deaths in some US cities \cite{Bootsma7588}.  The models presented here help to explain why this effect might occur, as illustrated in Fig.~\ref{fig:social_dist}.

The models presented here are certainly simplifications, making a variety of assumptions in order to increase understanding and to avoid over-fitting the limited data available; more complex models have been introduced and are currently in use \cite{Ferguson20COVID,Arenas}.  We note, though, that even between these rather simple models, the parameters obtained from our fits (Table \ref{fits}) can vary significantly for a given location and, though we have in each case determined which of these fits appears to have most validity, in many cases these are not strong indicators.  This variability illustrates the tremendous challenge of making accurate predictions of the course of the epidemic while still in its early stages and while operating under very limited data.  At the same time, this uncertainly may lend weight to the idea of erring on the side of caution, and continuing current social measures to curtail the pandemic.  Implementing such measures over a long period of time may prove prohibitively difficult, requiring the development of alternative approaches or policies that will allow more activities to proceed while continuing to reduce the spread of the virus.

\section*{Materials and Methods}

\subsubsection*{Relation between the exponential model and compartment models} 
The exponential model is appropriate during the first stages of the outbreak, when recoveries and deaths are negligible: in this case, the SIR compartment model can be directly reduced to an exponential model. If we assume $S\approx N$ in equations~\eqref{SIR}, then $dI(t)/dt\approx (\beta-\gamma)I$, with the exponential solution $I(t)=I_0e^{\alpha t}$ with $\alpha=\beta-\gamma$ and $I_0$ the initial number of infections.
We expect at very early times ($t\ll 1/\gamma$) that the recovery will lag infections so one might see $\alpha\sim\beta$ at very early times and then reduce to $\alpha\sim\beta-\gamma$ once $t> 1/\gamma$.  
Reports and graphs disseminated by the media typically report cumulative infections, which include recoveries and deaths. Using the SIR model, cumulative infections are $I_{c}(t)=I(t)+R(t)$ and evolve as $dI_{c}(t)/dt=\beta sI$.
Integrating this, we see that $I_{c}$ likewise grows exponentially with the same rate $\alpha=\beta-\gamma$.  
An important observation is that the doubling time for cumulative infections ($T_d=\ln(2)/\alpha$) will change during the early times, with a shorter doubling time while ($t\ll 1/\gamma$) and a longer doubling time when $t> 1/\gamma$. 

\subsubsection*{Relation between the HawkesN and SIR model}

Here we make the connection between the HawkesN process in Equation \ref{eq:SIRHawkes} and the SIR model in Equation \ref{SIR}.  Following \cite{SIRHawkes,yan2008distribution}, first a stochastic SIR model can be defined where a counting process 
$
C_t =N-S_t
$
tracks the total number of infections up to time $t$, $N$ is the population size, and $S_t$ is the number of susceptible individuals.  The process satisfies
\[
P(dC_t =1)=\beta S_t I_t dt/N+o(dt)
\]
\[
 P(dR_t =1)=\gamma I_t dt+o(dt)~,
\]
which then gives the rate of new infections and new recoveries as \cite{SIRHawkes}
\[
\lambda^I(t)=\beta S_t I_t/N,
\quad
\lambda^R(t)=\gamma I_t.
\]
It is shown in \cite{yan2008distribution} that the continuum limit of the counting process approaches the solution to the SIR model in Equation \ref{SIR}.  Furthermore, if the kernel $w(t)$ in the HawkesN model defined by \eqref{eq:SIRHawkes} is chosen to be exponential with parameter $\gamma$ and the reproduction number is chosen to be constant ($\R_0$), then 
$
E[\lambda^I(t)]=\lambda^H(t)
$
where $\mu=0$, $\beta=\R_0\gamma$ (see \cite{SIRHawkes} for further details).

\subsubsection*{Self-similar behavior of SIR}

 Calling the rescaled time $\tau=t\gamma$, \eqref{SIR} can be written as
   \begin{multline}
\frac{ds}{d\tau}=-\R_0 i s, \, 
\frac{di}{d\tau}=\R_0 is- i,  \,   \frac{dr}{d\tau}=\gamma i, \, \\ (s,i,r)|_{\tau=0} = (1-\epsilon, \epsilon, 0), \label{SIRsimilar}
\end{multline}
 where $0<\epsilon <<1$ is the initial fraction of the infected population at the start time. 

\subsubsection*{Fitting the SIR, SEIR, and Hawkes models}

The parameters in Table \ref{fits} were found using maximum Poisson likelihood regression \cite{Bootsma7588} via grid search with ranges $I_0\in[.005,.1]$, $\R_0\in[1.5,5]$, $\gamma \in [.01,.2]$, and $\mu \in [.01,.4]$.  Models were fit to empirical new infections per day or new mortality counts per day (assuming that 1\% of those labeled as Resistant in the simulations were the mortalities).  Fitted intensities are shown for each of the models in Table \ref{fits}.  We note that for the results presented here the likelihood is somewhat flat at the maximum, with multiple parameter combinations yielding plausible fits.  Fitting SIR type models is known to be challenging due to parameter identifiability issues \cite{evans2005structural,roosa2019assessing}.  The HawkesN process was fit using a non-parametric expectation maximization algorithm, the details of which can be found in \cite{HawkesR0COVID}.  The inter-infection time distribution is modeled using a Weibull distribution with shape $k$ and scale $b$.

\section*{Acknowledgments}
We thank Mark Lewis and Mason Porter for helpful comments on early drafts of the manuscript.  We thank Ron Brookmeyer for advice about the topic.  This research was supported by NSF grants DMS-1737770, SCC-1737585, ATD-1737996 and  the Simons Foundation Math + X Investigator award number 510776.


\bibliographystyle{plain}  
\bibliography{bib}

\begin{thebibliography}{10}

\bibitem{AIC}
H.~Akaike.
\newblock A new look at the statistical model identification.
\newblock {\em IEEE Transactions on Automatic Control}, 19(6):716--723, 1974.

\bibitem{Arenas}
Alex Arenas, Wesley Cot, Jesus Gomes-Gardenes, Sergio Gomez, Clara Granell,
  Joan~T. Matamalas, David Soriano, and Benjamin Steinegger.
\newblock A mathematical model for the spatiotemporal epidemic spreading of
  covid19.
\newblock 2020.
\newblock medRxiv article doi: https://doi.org/10.1101/2020.03.21.20040022.

\bibitem{Bootsma7588}
Martin C.~J. Bootsma and Neil~M. Ferguson.
\newblock The effect of public health measures on the 1918 influenza pandemic
  in {U.S.} cities.
\newblock {\em Proceedings of the National Academy of Sciences},
  104(18):7588--7593, 2007.

\bibitem{cauchemez2006real}
Simon Cauchemez, Pierre-Yves Bo{\"e}lle, Christl~A Donnelly, Neil~M Ferguson,
  Guy Thomas, Gabriel~M Leung, Anthony~J Hedley, Roy~M Anderson, and
  Alain-Jacques Valleron.
\newblock Real-time estimates in early detection of sars.
\newblock {\em Emerging infectious diseases}, 12(1):110, 2006.

\bibitem{cowling2010effective}
Benjamin~J Cowling, Max~SY Lau, Lai-Ming Ho, Shuk-Kwan Chuang, Thomas Tsang,
  Shao-Haei Liu, Pak-Yin Leung, Su-Vui Lo, and Eric~HY Lau.
\newblock The effective reproduction number of pandemic influenza: prospective
  estimation.
\newblock {\em Epidemiology (Cambridge, Mass.)}, 21(6):842, 2010.

\bibitem{plague}
Ensheng Dong, Hongru Du, and Lauren Gardner.
\newblock An interactive web-based dashboard to track {COVID-19} in real time.
\newblock {\em The Lancet}, 2020.
\newblock https://plague.com/.

\bibitem{evans2005structural}
Neil~D Evans, Lisa~J White, Michael~J Chapman, Keith~R Godfrey, and Michael~J
  Chappell.
\newblock The structural identifiability of the susceptible infected recovered
  model with seasonal forcing.
\newblock {\em Mathematical biosciences}, 194(2):175--197, 2005.

\bibitem{farrington2003branching}
CP~Farrington, MN~Kanaan, and NJ~Gay.
\newblock Branching process models for surveillance of infectious diseases
  controlled by mass vaccination.
\newblock {\em Biostatistics}, 4(2):279--295, 2003.

\bibitem{Ferguson20COVID}
Neil~M Ferguson, Daniel Laydon, Gemma Nedjati-Gilani, Natsuko Imai, Kylie
  Ainslie, Marc Baguelin, Sangeeta Bhatia, Adhiratha Boonyasiri, Zulma
  Cucunub\'a, Gina Cuomo-Dannenburg, Amy Dighe, Ilaria Dorigatti, Han Fu, Katy
  Gaythorpe, Will Green, Arran Hamlet, Wes Hinsley, Lucy~C Okell, Sabine van
  Elsland, Hayley Thompson, Robert Verity, Erik Volz, Haowei Wang, Yuanrong
  Wang, Patrick~GT Walker, Caroline Walters, Peter Winskill, Charles Whittaker,
  Christl~A Donnelly, Steven Riley, and Azra~C Ghani.
\newblock Impact of non-pharmaceutical interventions {(NPIs)} to reduce
  {COVID-19} mortality and healthcare demand.
\newblock 2020.
\newblock DOI: https://doi.org/10.25561/77482.

\bibitem{forsberg2008likelihood}
Laura Forsberg~White and Marcello Pagano.
\newblock A likelihood-based method for real-time estimation of the serial
  interval and reproductive number of an epidemic.
\newblock {\em Statistics in medicine}, 27(16):2999--3016, 2008.

\bibitem{Fox}
E.~W. Fox, M.~B. Short, F.~P. Schoenberg, K.~D. Coronges, , and A.~L. Bertozzi.
\newblock Modeling e-mail networks and inferring leadership using self-exciting
  point processes.
\newblock {\em J. Am. Stat. Assoc.}, 111(514):564--584, 2016.

\bibitem{Harko14}
Tiberiu Harko, Francisco S.~N. Lobo, and M.~K. Mak.
\newblock Exact analytical solutions of the susceptible-infected-recovered
  (sir) epidemic model and of the sir model with equal death and birth rates.
\newblock {\em Applied Mathematics and Computation}, 236:184--194, 2014.

\bibitem{EbolaHawkes}
R.~Harrigan, M.~Mossoko, E.~Okitolonda-Wemakoy, F.P. Schoenberg, N.~Hoff,
  P.~Mbala, S.R. Wannier, S.D. Lee, S.~Ahuka-Mundeke, T.B. Smith, B.~Selo,
  B.~Njokolo, G.~Rutherford, A.W. Rimoin, J.J.M. Tamfum, and J.~Park.
\newblock Real-time predictions of the 2018-2019 ebola virus disease outbreak
  in the democratic republic of congo using hawkes point process models.
\newblock {\em Epidemics}, 28, 2019.

\bibitem{hellewell2020feasibility}
Joel Hellewell, Sam Abbott, Amy Gimma, Nikos~I Bosse, Christopher~I Jarvis,
  Timothy~W Russell, James~D Munday, Adam~J Kucharski, W~John Edmunds, Fiona
  Sun, et~al.
\newblock Feasibility of controlling covid-19 outbreaks by isolation of cases
  and contacts.
\newblock {\em The Lancet Global Health}, 2020.

\bibitem{Imai}
Natsuko Imai, Anne Cori, Ilaria Dorigatti, Marc Baguelin, Christl~A. Donnelly,
  Steven Riley, and Neil~M. Ferguson.
\newblock Transmissibility of 2019-ncov.
\newblock 2020.
\newblock DOI: https://doi.org/10.25561/77148.

\bibitem{SIR27}
W.~O. Kermack and A.~G. McKendrick.
\newblock A contribution to the mathematical theory of epidemics",
  journal="proceedings of the royal society a.
\newblock 115(772):700--721, 1927.

\bibitem{EarlyR0}
Adam~J Kucharski, Timothy~W Russell, Charlie Diamond, Yang Liu, John Edmunds,
  Sebastian Funk, and Rosalind~M Eggo.
\newblock Early dynamics of transmission and control of covid-19: a
  mathematical modelling study.
\newblock {\em The Lancet, Infectious Diseases}, 2020.
\newblock March 11, 2020.

\bibitem{Lander20}
Mark Landler and Stephen Castle.
\newblock Behind the virus report that jarred the {U.S.} and the {U.K.} to
  action.
\newblock {\em The New York Times}, 2020.
\newblock March 17.

\bibitem{NEJMWuhan}
Q.~Li, X.~Guan, P.~Wu, X.~Wang, L.~Zhou, Y.~Tong, R.~Ren, K.S. Leung, E.H. Lau,
  J.Y. Wong, and 2020 Xing, X.
\newblock Early transmission dynamics in {Wuhan, China}, of novel coronavirus
  infected pneumonia.
\newblock {\em New England Journal of Medicine}, 2020.

\bibitem{li2020simulating}
Tianyi Li.
\newblock Simulating the spread of epidemics in china on the multi-layer
  transportation network: Beyond the coronavirus in wuhan, 2020.

\bibitem{LiuTravelMedicine2020}
Ying Liu, Albert~A Gayle, Annelies Wilder-Smith, and Joacim Rocklov.
\newblock The reproductive number of covid-19 is higher compared to sars
  coronavirus.
\newblock {\em Journal of Travel Medicine}, 27(2), 02 2020.
\newblock taaa021.

\bibitem{Rpackage}
Sebastian Meyer, Leonhard Held, and Michael {H\"ohle}.
\newblock Spatio-temporal analysis of epidemic phenomena using the {R} package
  surveillance.
\newblock {\em Journal of Statistical Software}, 77(11), 2017.

\bibitem{Miller12}
J.C. Miller.
\newblock A note on the derivation of epidemic final sizes.
\newblock {\em Bulletin of Mathematical Biology}, 74(9), 2012.
\newblock section 4.1.

\bibitem{Miller17}
J.C. Miller.
\newblock Mathematical models of sir disease spread with combined non-sexual
  and sexual transmission routes.
\newblock {\em Infectious Disease Modelling}, 2, 2017.
\newblock section 2.1.3.

\bibitem{HawkesR0COVID}
George Mohler, Frederic Schoenberg, Martin~B. Short, and Daniel Sledge.
\newblock Analyzing the world-wide impact of public health interventions on the
  transmission dynamics of covid-19.
\newblock 2020.
\newblock Researchgate DOI: 10.13140/RG.2.2.32817.12642.

\bibitem{Obadia12}
Thomas Obadia, Romana Haneef, and Pierre-Yves {Bo\"elle}.
\newblock The {$R_0$} package: a toolbox to estimate reproduction numbers for
  epidemic outbreaks.
\newblock {\em BMC medical informatics and decision making}, 12(1), 2012.

\bibitem{pan2020time}
Feng Pan, Tianhe Ye, Peng Sun, Shan Gui, Bo~Liang, Lingli Li, Dandan Zheng,
  Jiazheng Wang, Richard~L Hesketh, Lian Yang, et~al.
\newblock Time course of lung changes on chest ct during recovery from 2019
  novel coronavirus covid-19 pneumonia.
\newblock {\em Radiology}, page 200370, 2020.

\bibitem{Perkins}
T.~Alex Perkins, Sean~M. Cavany, Sean~M. Moore, Rachel~J. Oidtman, Anita Lerch,
  and Marya Poterek.
\newblock Estimating unobserved {SARS-CoV-2} infections in the united states.
\newblock 2020.
\newblock medRxiv article doi: https://doi.org/10.1101/2020.03.15.2003658.

\bibitem{riley2003transmission}
Steven Riley, Christophe Fraser, Christl~A Donnelly, Azra~C Ghani, Laith~J
  Abu-Raddad, Anthony~J Hedley, Gabriel~M Leung, Lai-Ming Ho, Tai-Hing Lam,
  Thuan~Q Thach, et~al.
\newblock Transmission dynamics of the etiological agent of sars in hong kong:
  impact of public health interventions.
\newblock {\em Science}, 300(5627):1961--1966, 2003.

\bibitem{Riou20}
Julien Riou and Christian~L. Althaus.
\newblock Pattern of early human-to-human transmission of wuhan 2019-ncov.
\newblock 2020.
\newblock bioRxiv article doi: https://doi.org/10.1101/2020.01.23.917351.

\bibitem{SIRHawkes}
Marian-Andrei Rizoiu, Swapnil Mishra, Quyu Kong, Mark Carman, and Lexing Xie.
\newblock {SIR-Hawkes}: Linking epidemic models and {Hawkes} processes to model
  diffusions in finite populations.
\newblock {\em Proc. of the 2018 {World Wide Web} Conference on {WWW}}, pages
  419--428, 2018.
\newblock https://arxiv.org/abs/1711.01679.

\bibitem{roosa2019assessing}
Kimberlyn Roosa and Gerardo Chowell.
\newblock Assessing parameter identifiability in compartmental dynamic models
  using a computational approach: application to infectious disease
  transmission models.
\newblock {\em Theoretical Biology and Medical Modelling}, 16(1):1, 2019.

\bibitem{RickAISM19}
F.P. Schoenberg, M.~Hoffmann, and R.~Harrigan.
\newblock A recursive point process model for infectious diseases.
\newblock {\em Annals of the Institute of Statistical Mathematics},
  71(5):1271--1287", 2019.

\bibitem{Stomakhin}
Alexey Stomakhin, Martin~B. Short, , and Andrea~L. Bertozzi.
\newblock Reconstruction of missing data in social networks based on temporal
  patterns of interactions.
\newblock {\em Inverse Problems}, 27(11):115013, 2011.

\bibitem{Tindale20}
Lauren~C. Tindale, Michelle Coombe, Jessica~E. Stockdale, Emma~S. Garlock, Wing
  Yin~Venus Lau, Manu Saraswat, Yen-Hsiang~Brian Lee, Louxin Zhang, Dongxuan
  Chen, Jacco Wallinga, and Caroline Colijn.
\newblock Transmission interval estimates suggest pre-symptomatic spread of
  covid-19.
\newblock 2020.
\newblock https://doi.org/10.1101/2020.03.03.20029983.

\bibitem{Verity}
Robert Verity et~al.
\newblock Estimates of the severity of {COVID-19} disease.
\newblock 2020.
\newblock medRxiv 2020.03.09.20033357; doi:
  https://doi.org/10.1101/2020.03.09.20033357.

\bibitem{wallinga2004different}
Jacco Wallinga and Peter Teunis.
\newblock Different epidemic curves for severe acute respiratory syndrome
  reveal similar impacts of control measures.
\newblock {\em American Journal of epidemiology}, 160(6):509--516, 2004.

\bibitem{Wiki-compartment}
Wikipedia.
\newblock Compartmental models in epidemiology, 2020.

\bibitem{WuLancet}
Joseph~T Wu, Kathy Leung, and Prof Gabriel~M Leung.
\newblock Nowcasting and forecasting the potential domestic and international
  spread of the 2019-ncov outbreak originating in wuhan, china: a modelling
  study.
\newblock {\em The Lancet}, 395(10225):698--697, 2020.
\newblock DOI:https://doi.org/10.1016/S0140-6736(20)30260-9.

\bibitem{yan2008distribution}
Ping Yan.
\newblock Distribution theory, stochastic processes and infectious disease
  modelling.
\newblock In {\em Mathematical epidemiology}, pages 229--293. Springer, 2008.

\bibitem{you2020estimation}
Chong You, Yuhao Deng, Wenjie Hu, Jiarui Sun, Qiushi Lin, Feng Zhou, Cheng~Heng
  Pang, Yuan Zhang, Zhengchao Chen, and Xiao-Hua Zhou.
\newblock Estimation of the time-varying reproduction number of {COVID-19}
  outbreak in china.
\newblock {\em Available at SSRN 3539694}, 2020.

\bibitem{zhou2020clinical}
Fei Zhou, Ting Yu, Ronghui Du, Guohui Fan, Ying Liu, Zhibo Liu, Jie Xiang,
  Yeming Wang, Bin Song, Xiaoying Gu, et~al.
\newblock Clinical course and risk factors for mortality of adult inpatients
  with covid-19 in wuhan, china: a retrospective cohort study.
\newblock {\em The Lancet}, 2020.

\bibitem{Zipkin}
Joseph~R. Zipkin, Frederic~P. Schoenberg, Kathryn Coronges, and Andrea~L.
  Bertozzi.
\newblock Point-process models of social network interactions: parameter
  estimation and missing data recovery.
\newblock {\em Eur. J. Appl. Math.}, 27(3):502--529, 2016.

\end{thebibliography}

\end{document}